\documentstyle[12pt]{article}
\def\ltap{\raisebox{-.4ex}{\rlap{$\sim$}} \raisebox{.4ex}{$<$}}
\def\gtap{\raisebox{-.4ex}{\rlap{$\sim$}} \raisebox{.4ex}{$>$}}
\def\journal{
        \topmargin 0in \oddsidemargin .0in
        \headheight 0pt \headsep 0pt
        \textwidth 6.5in
\textheight  9in
        \marginparwidth 1in
        \parindent 2em
        \parskip .5ex plus .1ex         \jot = 1.5ex}
\journal
\def\hm{Higgs mechanism\hspace {.05in}}
\def\lag5{\mbox{${\cal L}_{\rm SB}\hbox{ }$}}
\def\ltap{\raisebox{-.4ex}{\rlap{$\sim$}} \raisebox{.4ex}{$<$}}
\def\gtap{\raisebox{-.4ex}{\rlap{$\sim$}} \raisebox{.4ex}{$>$}}

\def\wpwp{$W^+W^+\hbox{ }$}

\def\u1{$U(1)\hbox{ }$}

\def\su2xu1{$SU(2)_L \times U(1)_Y\hbox{ }$}
\def\su2{$SU(2)\hbox{ }$}
\def\rpp{$\rho \pi \pi\hbox{ }$}
\def\rhotc{$\rho_{TC}\hbox{ }$}
\def\ro{$\rho\hbox{ }$}
\def\qrho{``$\rho$''\hbox{ }}

\def\qqb{$\overline{q}q \hbox{ }$}

\def\wlwl{$W_LW_L\hbox{ }$}
\def\tl{\vec{T}_L}
\def\wpwm{$W^+W^-\hbox{ }$}
\def\m5{$M_{\rm SB}\hbox{ }$}
\def\caM{{\cal{M}}}

\def\alphaw{$\alpha_W\hbox{ }$}
\def\alphaw2{$\alpha_W^2\hbox{ }$}
\def\alphawalphas{$\alpha_W \alpha_S\hbox{ }$}
\def\oalphaw2{$O(\alpha_W^2)\hbox{ }$}

\def\pp{$\pi \pi\hbox{ }$}

\def\ra{\rightarrow}
\input{epsf}
\begin{document}
\begin{titlepage}
\begin{center}
November 24, 1998                    \hfill    LBNL-42570\\

\vskip .5in

{\large \bf Strong $WW$ scattering at the end of the 90's:\\ theory and 
experimental prospects }\footnote
{This work was supported by the Director, Office of Energy
Research, Office of High Energy and Nuclear Physics, Division of High
Energy Physics of the U.S. Department of Energy under Contract
DE-AC03-76SF00098.}

\vskip .5in

Michael S. Chanowitz\footnote{Email: chanowitz@lbl.gov}

\vskip .2in
{\em Theoretical Physics Group\\
     Lawrence Berkeley National Laboratory\\
     University of California\\
     Berkeley, California 94720}
\end{center}

\vskip .25in 
\begin{abstract} 

The nature of electroweak symmetry breaking can only be established 
definitively by the direct discovery and detailed study of the 
symmetry breaking quanta at high energy colliders.  At the LHC the 
ability to observe TeV scale strong $WW$ scattering confers a no-lose 
capability to establish the mass scale and interaction strength of the 
symmetry breaking quanta, even if the symmetry breaking quanta resist 
discovery and whether strong $WW$ scattering is observed or excluded.  
This lecture discusses the motivation to consider strong $WW$ 
scattering in light of what we have learned from precision electroweak 
data during the decade.  The theoretical basis for strong $WW$ 
scattering is explained with an introductory review of the Higgs 
mechanism from a general perspective that encompasses light, 
perturbative Higgs bosons or nonperturbative, dynamical symmetry 
breaking by TeV scale strong interactions.  The experimental signals 
and backgrounds are reviewed and the sensitivity of experiments at the 
LHC is assessed.

\end{abstract} 
\vskip .5in
\begin{center}
\noindent {\em A lecture presented at the Zuoz summer school on Hidden 
Symmetries and Higgs Phenomena, August 17 -21, 1998, Zuoz, 
Switzerland, to be published in the proceedings.}
\end{center}

\end{titlepage}

\renewcommand{\thepage}{\roman{page}}
\setcounter{page}{2}
\mbox{ }

\vskip 1in

\begin{center}
{\bf Disclaimer}
\end{center}

\vskip .2in

\begin{scriptsize}
\begin{quotation}
This document was prepared as an account of work sponsored by the United
States Government. While this document is believed to contain correct
 information, neither the United States Government nor any agency
thereof, nor The Regents of the University of California, nor any of their
employees, makes any warranty, express or implied, or assumes any legal
liability or responsibility for the accuracy, completeness, or usefulness
of any information, apparatus, product, or process disclosed, or represents
that its use would not infringe privately owned rights.  Reference herein
to any specific commercial products process, or service by its trade name,
trademark, manufacturer, or otherwise, does not necessarily constitute or
imply its endorsement, recommendation, or favoring by the United States
Government or any agency thereof, or The Regents of the University of
California.  The views and opinions of authors expressed herein do not
necessarily state or reflect those of the United States Government or any
agency thereof, or The Regents of the University of California.
\end{quotation}
\end{scriptsize}

\vskip 2in

\begin{center}
\begin{small}
{\it Lawrence Berkeley National Laboratory is an equal opportunity employer.}
\end{small}
\end{center}

\newpage

\renewcommand{\thepage}{\arabic{page}}
\setcounter{page}{1}

\noindent {\bf 1.  Introduction} 

Broadly speaking there are two possibilities for electroweak symmetry 
breaking: weakly coupled Higgs bosons below 1 TeV or a new sector of 
quanta at the TeV scale that interact strongly with one another {\em 
and} with longitudinally polarized $W$ and $Z$ bosons.\footnote{In 
this sense, theories with TeV scale strong dynamics that engenders
a light composite Higgs scalar which breaks $SU(2) \times U(1)$ are 
classified as weakly coupled.} While precision electroweak data 
accumulated in the 90's favor the first scenario, the conclusion is 
not definitve.  The study of $WW$ scattering at the TeV scale, to 
begin at the LHC, will provide a no-lose capability to determine the 
strength and mass scale of the symmetry breaking quanta.  As discussed 
below, it is a fundamental measurement, of interest even if a light 
Higgs boson candidate were to be discovered before inauguration of the 
LHC, and it could be crucial if Higgs sector quanta turn out to be 
more elusive.  This lecture is an introduction to the motivation, 
theory, and techniques for the study of $WW$ scattering at the TeV 
scale.  In particular I will focus on the generic strong $WW$ 
scattering signal\cite{mcmkg2} that can be used at the LHC to 
determine definitively whether the symmetry breaking force is weak or 
strong.  Because we would learn from the presence or absence of strong 
$WW$ scattering whether the symmetry breaking physics is at or below 
the TeV scale, ability to measure or exclude the signal confers a 
``no-lose'' capability to determine the mass scale of symmetry 
breaking.

While the precision electroweak data favors the weak breaking 
scenario, the strong $WW$ scattering measurement remains an important 
tool for the study of symmetry breaking: 

First, because the precision electroweak data probes the Higgs quanta 
only indirectly, by means of quantum corrections, it can never 
definitively determine the nature of the Higgs sector.  The relevant 
quantum corrections are open to contributions from many forms of new 
physics.  Occam's razor favors the simplest 
interpretation, which assumes that the only new physics contributing 
significantly to the radiative corrections is the quanta 
that form the symmetry breaking condensate. In that case the data do 
favor weak symmetry breaking.  But nature may well have dealt us a 
more complicated hand, with new physics accompanying the symmetry 
breaking quanta also contributing to the radiative corrections.  Then 
the precision data tells us nothing about the symmetry breaking sector 
--- unless we can ``unscramble'' the different contributions, which in 
general we do not know how to do.  While my focus is on general 
aspects and not on specific models, reference \cite{strongmodels} 
lists a few models of strong symmetry breaking that can serve as 
existence proofs.  The nature of the symmetry breaking sector can only 
be established definitively by its direct discovery and detailed study 
in experiments at high energy colliders.  In the meantime it is 
sensible to be guided by compelling theories, e.g., SUSY, but not to 
rely on them exclusively.

Second, even if there were a light Higgs boson it is possible for it 
to be undetectable at all planned experiments up to and including the 
LHC. For example, Gunion, Haber, and Moroi\cite{ghm} have found 
``blind spots'' in the parameter space of the NMSSM (the 
next-to-minimal model, containing a Higgs singlet field in addition to 
the fields of the MSSM) for which none of the Higgs scalars could be 
discovered at LEP or the LHC, even if the LHC were to accumulate the 
heroic integrated luminosity of 600 fb$^{-1}$.  In that case the 
``no-lose'' capability of the LHC would be crucial to establish 
whether electroweak symmetry breaking is weak or strong.  As 
discussed in this lecture, with 
$\simeq$ 100 - 150 fb$^{-1}$ the LHC could observe or exclude strong 
$WW$ scattering.\cite{mcwk1,mcwk2,jbetal} In a blind-spot 
scenario it could establish the absence of strong $WW$ scattering 
which would tell us to look harder for light Higgs scalars below 1 
TeV, perhaps with an electron collider.\cite{ghm2}  Or,
if strong $WW$ scattering were observed, it would tell us to look 
for the Higgs sector quanta above 1 TeV, perhaps with a VLHC.

Third, even if there were a light Higgs boson and even if it were 
discovered at LEP, Fermilab, or the LHC, it would still be important 
to measure the $WW$ scattering cross section in the TeV region.  If 
symmetry breaking is due to a light Higgs boson, a central prediction 
of the Higgs mechanism is that there be no strong $WW$ 
scattering.  As explained in the review of the Higgs mechanism given 
below, strong $WW$ scattering is the first-cousin to the famous ``bad 
high energy behavior'' which it is the principle mission of the Higgs 
mechanism to remove.  For strong symmetry breaking the unitarity of 
longitudinal $WW$ scattering is saturated, while for weak breaking, 
longitudinal $WW$ scattering cuts off while it is still weak, well 
below where unitarity would be saturated.  In discussing the 
experimental signals at the LHC I will consider both the capability to 
observe strong $WW$ scattering if it is present and to exclude it if 
it is not.

The central point is easy to grasp: we have already discovered three 
quanta from the Higgs sector, the longitudinal spin modes of the 
$W^{\pm}$ and $Z$ bosons, which in the Higgs mechanism are the 
transubstantiated ghosts of Higgs sector quanta.  By measuring the 
scattering of the longitudinal modes $W_{L}W_{L} \rightarrow 
W_{L}W_{L}$ (where $L$ denotes longitudinal) we are probing the Higgs 
sector interactions, a statement made precise by the `equivalence 
theorem.'\cite {et1,et2} As reviewed below the absence or presence of 
an enhanced $WW$ signal at the LHC can then determine if the Higgs 
sector interactions are weak or strong and correspondingly if the 
Higgs sector quanta lie below or at the TeV scale.

The lecture is organized as follows:
\begin{itemize}

\item Section 2 reviews the Higgs mechanism in a general framework 
that applies whether Higgs bosons exist or not, using symmetry and 
unitarity to analyze the possible forms electroweak symmetry breaking 
may take.

\item Section 3 is a brief discussion of the implications of the 
precision electroweak data presented at the recent Vancouver ICHEP 
meeting.

\item Section 4 reviews models used to estimate strong $WW$ scattering 
cross sections at high energy colliders. 

\item Section 5 discusses methods of computing $WW$ scattering at 
colliders, including the `classical' effective $W$ 
approximation\cite{ewa} (EWA) and a more complete method, the 
effective Higgs bosons representation\cite{ehb} (EHB),  which predicts 
the experimentally important transverse momentum spectrum 
of the final state jets recoiling against the $WW$ pair that cannot 
be obtained from the EWA.

\item Section 6 considers the question `Can LHC lose?', reviewing the 
experimental strategies and the capability of LHC experiments to 
observe or exclude strong $WW$ scattering.

\item A brief conclusion is given in section 7.
\end{itemize}
\newpage

\noindent {\bf 2. General Framework} 

We begin with a general description of the Higgs mechanism in 
Section 2.1 that applies whether Higgs bosons exist or not.  
Implications for strong $WW$ scattering are reviewed in the subsequent 
subsections: the Equivalence Theorem in 2.2, the $WW$ low energy 
theorems in 2.3, and unitarity and the energy scale of strong $WW$ 
scattering in 2.4. 

\noindent {\em 2.1 The Generic Higgs Mechanism\cite{generic}}

The basic ingredients of the \hm are a gauge sector and a symmetry 
breaking sector, 
$$
{\cal L} = {\cal L}_{\hbox{\small{gauge}}} + \lag5.
\eqno(2.1)
$$
${\cal L}_{\hbox{\small{gauge}}}$ is an unbroken {\it locally 
symmetric = gauge invariant} theory, describing massless gauge bosons 
that are transversely polarized, just like the photon.  For $SU(2)_L 
\times U(1)_Y$ gauge symmetry the gauge bosons are a triplet $\vec{W} 
= W_1, W_2, W_3$ corresponding to the generators $\tl$ and a singlet 
gauge boson $X$ corresponding to the hypercharge generator $Y$.  \lag5 
is the symmetry breaking Lagrangian that describes the dynamics of the 
symmetry breaking force and the associated quanta.  If \lag5 did not 
exist, the unbroken $SU(2)_L$ nonabelian symmetry would give rise to a 
force that would confine quanta of nonvanishing $\tl$ charge, such as 
left-handed electrons and neutrinos.

In the generic Higgs mechanism $\lag5$ breaks the {\it local} (or {\it gauge})
 symmetry
of ${\cal L}_{\hbox{\small{gauge}}}$.  To do so 
$\lag5$ must possess a {\it global} symmetry
$G$ that breaks spontaneously to a subgroup $H$,
$$G \to H.
\eqno(2.2)$$
In the electroweak theory we do not yet know either of the groups $G$ or $H$,
$$G = \ ? \eqno(2.3a)$$
$$H = \ ? \eqno(2.3b)$$
We want to discover what they are and beyond that we want to discover the
symmetry breaking sector
$$\lag5 = \ ? \eqno(2.4)$$
including the mass scale of its spectrum
$$M_{\rm SB} = \ ? \eqno(2.5)$$
and the interaction strength
$$\lambda_{\rm SB} = \ ? \eqno(2.6)$$

We do already know one fact about $G$ and $H$:
$G$ must be at least as big as $SU(2)_L \times U(1)_Y$ or $\lag5$ would {\it
explicitly} (as opposed to {\it spontaneously}) break the
$SU(2)_L \times U(1)_Y$
gauge symmetry.  Similarly $H$ must be at least as big as $U(1)_{EM}$ or the
theory after spontaneous breakdown will not accommodate the unbroken gauge
symmetry of QED.  That is, in order to be consistent with the desired pattern of
breaking for the {\it local} symmetry
$$SU(2)_L \times U(1)_Y \to U(1)_{EM} \eqno(2.7)$$
the spontaneous breaking of the {\it global} symmetry of $\lag5$
$$G \to H \eqno(2.8)$$
is constrained by
$$
G \supset SU(2)_L \times U(1)_Y  (2.9)
$$
$$
H \supset  U(1)_{EM}.  (2.10)
$$

There are two steps in the Higgs mechanism.  The first has nothing to do with
gauge symmetry---it is just the spontaneous breaking of a global symmetry as
explained by the Goldstone theorem. By {\it spontaneous} symmetry
breaking $G \to H$ we mean that
$$G = \mbox{ global symmetry of the {\it interactions} of }\lag5
\eqno(2.11a)$$
while
$$H = \mbox{ global symmetry of the {\it ground-state} of } \lag5.
\eqno(2.11b)$$
That is, the dynamics of $\lag5$ are such that the state of lowest energy (the
{\it vacuum} in quantum field theory) has a smaller symmetry group than the
symmetry of the force laws of the Lagrangian.
Goldstone's theorem tells us that for each
broken generator of $G$ the spectrum of $\lag5$ contains a massless spin zero
particle or Goldstone boson,
$$
\mbox{\# of }  \mbox{massless scalars}
              = \mbox{ \# of broken symmetry axes}
             = \mbox{ dimension } G - \mbox{dimension H}$$
$$               = \mbox{ \# of energetically flat directions in field
space.} \eqno(2.12)$$
The last line is the clue to the proof of the theorem: masses arise from terms
that are quadratic in the fields,
$${\cal L}_{\hbox{mass}} = - {1\over 2} m^2 \phi^2,
\eqno(2.13)$$
so a field direction that is locally flat in energy (i.e., goes like $\phi^n$
with $n \geq 3$) corresponds to a massless mode.

The essential features are the symmetries of the Lagrangian $(G)$ and the
ground state $(H)$.
Elementary scalars are {\it not} essential:
if necessary Nature will make composite massless scalars.
She has (almost) already done so on at least one occasion: 
we believe on the basis of strong theoretical and experimental
evidence that QCD with two (almost) massless quarks is an example, with the
pion isotriplet the (almost) Goldstone bosons.
The initial global (flavor) symmetry of two flavor QCD in the $m_u = m_d =0$
limit is
$$
G=SU (2)_L\times SU(2)_R\eqno(2.14)
$$
since we could perform separate isospin  rotations on  
the right and left chirality $u$ and $d$ quarks.
The ground state has a nonvanishing expectation value for the
bilinear operator
$$
\langle \overline{u}_Lu_R +\overline{d}_Ld_R+h.c.\rangle_0\neq 0\eqno(2.15)
$$
where $h.c.=$ hermitian conjugate.
The condensate (2.15) breaks the global symmetry spontaneously, $G\to H$, where
$$
H=SU(2)_{L+R}\eqno(2.16)
$$
is the ordinary isospin group of nuclear and hadron physics.  That is, 
(2.15) is not invariant under independent rotations of left and right 
helicity quarks but only under rotations that act equally on left and 
right helicities.  In this example dim $G=6$ and dim $H=3$ so we 
expect $6-3=3$ Goldstone bosons.  In nature we believe they are the 
pion triplet, $\pi^+,\pi^-,\pi^0$, which are much lighter than typical 
hadrons because the $u$ and $d$ quark masses are so small on the 
hadronic scale, of order 10 MeV. (I refer to the ``current'' quark 
masses, the parameters that appear in the QCD Lagrangian.)

In the first step we considered only the global symmetry breakdown 
induced by $\lag5$ --- Goldstone's theorem.  Now we come to the second 
step, which involves the interplay of $\lag5$ with 
${\cal{L}}_{\hbox{\small{gauge}}}$.  The essential point of the Higgs 
mechanism is that when a spontaneously broken generator of $\lag5$ 
coincides with a generator of a gauge invariance of 
${\cal{L}}_{\hbox{\small{gauge}}}$, the associate Goldstone boson $w$ 
and massless gauge boson $W$ mix to form a massive gauge boson.  The 
number of degrees of freedom are preserved, since the Goldstone boson 
disappears from the physical spectrum while the gauge boson acquires a 
third (longitudinal) polarization state.  Like the first step this is 
a general phenomenon that depends only on the nature of the global and 
local symmetries, regardless of whether there are elementary scalar 
particles in the theory.

Suppose the Goldstone boson $w$ couples to one of the gauge currents,
with a coupling strength $f$ which has the dimension of a mass,
$$
\langle 0\vert J^\mu_{\hbox{\small{gauge}}} \vert w (p)\rangle = {i\over 2}
fp^\mu\eqno(2.17)
$$
$f$ is analogous to $F_\pi$, the pion decay constant, that specifies the
coupling of the pion to the axial isospin current,
$$
\langle 0\vert J^\mu_5 \vert \pi (p)\rangle = i
F_{\pi}p^\mu\eqno(2.18)
$$
Equation (2.17) means that the current contains
a term linear in $w$,
$$
J^\mu_{\hbox{\small{gauge}}} (x) ={1\over 2} f \partial^\mu w(x)+\cdots\eqno(2.18)
$$
In the Lagrangian $J^\mu_{\hbox{\small{gauge}}}$ is by definition coupled 
to the gauge boson
$W^\mu$,
$$
{\cal{L}}_{\hbox{\small{gauge}}} = g W_\mu J^\mu_{\hbox{\small{gauge}}}
+\cdots\eqno(2.19)
$$
where $g$ is the dimensionless gauge coupling constant.
Substituting Eq. (2.17) we find
$$
{\cal{L}}_{\hbox{\small{gauge}}}={1\over 2} gf W_\mu
(\partial^\mu w)\ldots\eqno(2.20)
$$
which shows that $W_\mu$ mixes in the longitudinal (parallel to $\vec{p}$)
direction with the would--be Goldstone boson $w$.

We can use (2.20) to compute the $W$ mass.\cite{generic} 
In the absence of symmetry breaking the $W$ is massless and transversely 
polarized. Therefore as in QED we can write its  propagator in Landau gauge,
$$
D^{\mu\nu}_0 = {-i\over k^2} (g^{\mu\nu}-{k^\mu k^\nu\over k^2})
\eqno(2.21)
$$
In higher orders the propagator is the sum of the geometric series due to
``vacuum polarization'',
i.e., all states that mix with the gauge current.
The vacuum polarization tensor is defined as
$$
\Pi^{\mu\nu}(k) =- \int d^4 k e^{-ik\cdot x}\langle TJ^\mu (x) J^\nu
(0)\rangle_0\\
= i{g^2f^2\over 4} (g^{\mu\nu}-{k^\mu k^\nu\over k^2})+\cdots 
\eqno(2.22)
$$
In Eq. (2.22) I have indicated explicitly the 
contribution from the Goldstone boson
pole: the factor $1/k^2$ is just the massless propagator and the factor
$(gf/2)^2$ can be recognized from Eq. (2.20).
The $g^{\mu\nu}$ term 
is present since gauge invariance requires current conservation,
$k_\mu \Pi^{\mu\nu}=0$.
Since it is a constant term with no absorptive part, it does not
affect the spectrum of the theory.
(In theories with elementary scalars it arises automatically from the
contact interaction given by the Feynman rules.)

Finally we compute the $W$ propagator from the geometric series
$$
D^{\mu\nu}  = (D_0 + D_0\Pi D_0 +\ldots )^{\mu\nu}
           = - {i\over k^2} \left(g^{\mu\nu}-{k^\mu k^\nu\over k^2}\right)
                \left(1+ {g^2 f^2\over 4k^2}+\cdots \right)
$$
$$                
           = -i \left(g^{\mu\nu} - {k^\mu k^\nu \over k^2}\right){1\over k^2}
                {1\over 1-\displaystyle{{g^2f^2\over 4k^2}}}
           = -i {g^{\mu\nu}-{\displaystyle{k^\mu k^\nu\over k^2}}\over k^2-
{\displaystyle{g^2f^2\over 4}}}.
           \eqno (2.23) 
$$
The massless Goldstone boson pole then induces a shift in the 
pole of the gauge boson propagator(!), to 
$$
M_W ={1\over 2} gf.\eqno(2.24)
$$
From the measured value of the Fermi constant,
$$
G_F={g^2\over 4\sqrt{2}M^2_W}={1\over \sqrt{2}f^2}\eqno(2.25)
$$
we learn that
$$
f\simeq 250\ {\rm GeV.} \eqno(2.26)
$$

Customarily instead of $f$ we refer to $v\equiv f$, the so--called vacuum
expectation value.
This custom, which I will also follow (though in general it is not really 
correct) derives from theories with elementary scalar fields where
$v\equiv f$ is both the coupling strength of the Goldstone boson $w$ to
$J_{\hbox{\small{gauge}}}$, as in (2.17), and  is also the value of the Higgs
boson field in the ground state (i.e., the Higgs boson vacuum condensate).
However the derivation just reviewed shows  that there is no need for a
Higgs boson to exist.
The condensate that  breaks the symmetry may be that of a composite
operator, e.g., Eq. (2.15), which in general has no simple relationship to the
parameter $f\equiv v$ defined in (2.17).
For instance, in QCD there is no trivial relationship between $F_\pi$ and
$\langle \overline{u}u +\overline{d}d \rangle_0$ (although there is a nontrivial
relation involving also the quark and pion masses).

\noindent {\em 2.2 The Equivalence Theorem}

The equivalence theorem\cite{et1,et2}
 is very useful for analyzing the implications 
of the Higgs mechanism for strong $WW$ scattering.
In the $U$ (unitary) gauge the Goldstone boson fields $\vec{w}$ are absent from
the Lagrangian. In $R$ (renormalizable) gauges they
do appear in \lag5 and in the Feynman rules,
though gauge invariance ensures that are not in the physical spectrum.
Since they engender the longitudinal gauge boson modes, $W_L$ and $Z_L$,
it is plausible that $W_L$ and $Z_L$ interactions reflect the dynamics of
$\vec{w}$.
The equivalence theorem is the precise statement of this proposition,
$$
{\cal{M}}(W_L(p_1), W_L(P_2), \ldots ) = {\cal{M}} (w(p_1), w(p_2),\ldots )_R
+O\left({M_W\over E_i}\right).\eqno(2.27)
$$
As indicated the equality holds up to corrections of order $M_W/E_i$.

We will see that the equivalence theorem is useful in the derivation of the
$W_LW_L$ low energy theorems and that it is also a useful source of intuition
for the possible dynamics of strong $WW$ scattering. In addition, it
greatly simplifies perturbative computations.
For instance, the evaluation of heavy Higgs boson production via $WW$ fusion
in unitary gauge requires evaluation of many diagrams with
``bad'' high energy behavior that cancel to give the final result.
But to leading order in the strong coupling $\lambda = m^2_H /2v^2$ it
suffices using the equivalence theorem to compute just a few simple diagrams.
The result embodies the cancellations of many diagrams in unitary gauge
and trivially has the correct high energy behavior. It is very accurate
for energies above 1 TeV (of order 1 \% or better).

A simple example is instructive.  Consider the decay of a heavy 
Higgs boson to a pair of longitudinally polarized gauge bosons 
$W^+_LW^-_L$.  In unitary gauge the $H W^+_LW^-_L$ amplitude is
$$
{\cal{M}} (H\to W^+_LW^-_L ) = g M_W \epsilon_L (p_1)\cdot \epsilon_L
(p_2).\eqno(2.28)
$$
For $m_H \gg M_W$ we neglect terms of order $M_W /m_H$, so that
$\epsilon^\mu_L  (p_i) \cong p_i/M_W$ and similarly from
$m^2_H = (p_1 + p_2)^2 \cong 2p_1\cdot p_2$ we find
$$
{\cal{M}} (H\to W^+_LW^-_L )= g{m^2_H\over 2M_W}+O\left({M_W\over
m_H}\right).\eqno(2.29)
$$
In a renormalizable gauge the corresponding amplitude can be read off
(taking care with factors of 2) from the $Hww$ vertex  in the Higgs potential,
with the result
$$
{\cal{M}} (H\to w^+w^- )= 2\lambda v.\eqno(2.30)
$$

Using the relations $M_W ={1\over 2} gv$ and $\lambda = m_H^2/2v^2 $ 
we see that Eqs.  (2.29) and (2.30) are indeed equal up to 
$O(M_W/m_H)$ corrections.

The theorem was first proved in tree approximation\cite{et1} and used in 
a variety of calculations. (Lee {\it et al}.\cite{et1}  contains a proof 
to all orders which does not however apply to matrix elements with 
more than one external $W_L$.)  Proofs to all orders in both $\lag5$ 
and ${\cal{L}}_{\hbox{\small{gauge}}}$ are given in \cite{et2}.
The fact that the theorem holds to all orders in 
the strong interactions of \lag5 is crucial for its applicability to 
strong $WW$ scattering.  

\noindent {\it 2.3 Low Energy Theorems}

Using the equivalence theorem and the general properties of the Higgs 
mechanism described in Section 2.1 we can derive the low energy 
theorems for \wlwl scattering that in turn set the scale for the onset 
of strong $WW$ scattering.  The symmetry breaking pattern of \lag5, $G 
\rightarrow H$, implies low energy theorems for the Goldstone bosons, 
which imply \wlwl low energy theorems by means of the equivalence 
theorem.  In general the low energy theorems are determined by the 
groups $G$ and $H$ and by two parameters, the vacuum expectation value 
$v$ and the $\rho$ parameter, $\rho = M_W^2/(M_Z^2\cos^2\theta_W)$.  
Recall that we assume that there are no light quanta in \lag5 other 
than $w$ and $z$.  If there are other light quanta in \lag5 they may 
or may not modify the low energy theorems.\cite{let}

Low energy theorems for the $2 \to 2$ scattering of Goldstone bosons were first
derived by Weinberg\cite{wbglet} for pion-pion scattering.
Identifying the pion isotriplet with the almost-Goldstone bosons of
spontaneous flavor symmetry breaking 
$SU(2)_L\times SU(2)_R\to SU(2)_{L+R}$ in hadron physics, Weinberg showed for
example that
$$
{\cal{M}}(\pi^+\pi^- \to \pi^0\pi^0 )= {s\over F^2_\pi}\eqno(2.31)
$$
where $F_\pi = 93$ MeV is the pion decay constant defined in Eq. (2.18).
Equation (2.31) neglects $O(m^2_\pi)$ corrections (which are in fact calculable
to leading order and were computed by Weinberg) 
and is valid at low energy, defined as
$$
s \ll\ \hbox{minimum}\{ m^2_\rho , (4\pi F_\pi)^2
\}.\eqno(2.32)
$$

The low energy theorems can be derived by current algebra or effective
Lagrangian methods.
The proof has two important features:
\begin{itemize}
\item it is valid to all orders in the Goldstone boson self--interactions.
This is crucial since those interactions may be strong (as they are for the
pion example) so that perturbation theory is a non-starter,
\item we needn't be able to solve the dynamics or even to know the Lagrangian
of the theory.
In fact the $\pi\pi$ low energy theorems were derived in 1966 before QCD was
discovered. (And we still don't know today how to compute $\pi\pi$ scattering
directly in QCD.)
\end{itemize}
The current algebra/symmetry method was important in the path followed in the
1960's that led in the early 1970's to the discovery that
${\cal{L}}_{{HADRON}} = {\cal{L}}_{QCD}$. It is similarly useful today in our
search for $\lag5$.

If $G= SU(2)_L\times SU(2)_R$ and $H= SU(2)_{L+R}$ as in QCD, then we can
immediately conclude, just as in Eq. (2.31) that$^1$
$$
{\cal{M}} (w^+w^-\to zz)={s\over v^2}\eqno(2.33)
$$
at low energy,
$$
s \ll\ \hbox{minimum}\{M^2_{\rm SB}, (4\pi v)^2\},\eqno(2.34)
$$
as in eq. (2.32).
Here $M_{\rm SB}$ is the typical mass scale of $\lag5$ 
and $v\simeq {1\over 4}$ TeV.
More generally,
electroweak gauge invariance requires Eqs. (2.9) and (2.10) from which we can 
deduce the more general result\cite{let}
$$
{\cal{M}}(w^+w^- \to zz) = {1\over \rho}{s\over v^2}.\eqno(2.35)
$$
Equation (2.35) is arguably more soundly based than (2.31) was in 1966, since
(2.35) is a general consequence of gauge invariance and the Higgs mechanism
while (2.31) was based on inspired guesswork as to the symmetries underlying
hadron physics. The low energy theorems are proved by three different
methods\cite{let}: perturbatively, by a current algebra
derivation similar to Weinberg's, and by the chiral Lagrangian method.

We can next use the equivalence theorem, Eq. (2.27), to turn Eq. (2.35) into a
physical statement about longitudinal gauge boson scattering.
In  particular we have
$$
{\cal{M}}(W_L^+W_L^-\to Z_LZ_L)={1\over\rho}{s\over v^2}\eqno(2.36)
$$
for an energy domain circumscribed by Eqs. (2.34) and (2.27) as
$$
M^2_W \ll s \ll \hbox{minimum}\{M^2_{\rm SB}, (4\pi v)^2\}.\eqno(2.37)
$$
The window (2.37) may or may not exist in nature, depending on 
whether $M_{\rm SB}
\gg M_W$.

In addition to Eq. (2.36) there are two other independent amplitudes which may
be chosen to be \wpwm and $ZZ$ elastic scattering. Their low energy theorems
are:
$$
\caM (W^+_LW^-_L \to W^+_LW^-_L)=-\left(4-{3\over\rho} \right){u\over 
v^2},\eqno(2.38)$$
$$
\caM (Z_LZ_L \to Z_LZ_L ) = 0.\eqno(2.39)
$$
There are in addition four others that can be obtained by crossing symmetry:
$$
\caM (W_L^\pm Z_L \to W_L^\pm Z_L )= {1\over \rho}{t\over v^2}, \eqno(2.40)
$$
$$
\caM (W_L^+W^+_L \to W^+_LW^+_L)
=\caM (W^-_L W^-_L \to W^-_L W^-_L ) =- (4 - {3\over \rho}){s\over 
v^2}.\eqno (2.41)
$$

\noindent {\it 2.4 Unitarity and the Scale of Strong $WW$ Scattering}

The threshold energy dependence predicted by the low energy theorems 
would eventually violate unitarity unless damped.  In fact, the low 
energy theorems are identical with the famous ``bad'' high energy 
behavior that the Higgs mechanism was invented to cure --- this 
emerges most clearly in the perturbative derivation.$^3$ Within the 
\hm it is the task of $\lag5$ to cut off the growing amplitudes Eqs.  
(2.36-2.41).  Unitarity implies a rigorous upper bound on the energy 
at which this must occur.  The use of unitarity here is identical to 
that of Lee and Yang\cite{leeyang} and of Ioffe, Okun, and 
Rudik\cite{ior} who used the growing behavior of fermion-fermion 
scattering in Fermi's four-fermion weak interaction Lagrangian (also 
proportional to $G_{F}s \propto s/v^2$ !)  to bound the scale at which 
Fermi's theory must break down --- essentially a bound on the mass of 
the $W$ boson.  In fact that bound is precisely half the value of the 
bound given below for the scale of the symmetry breaking physics.

In particular we use partial wave unitarity.
The partial wave amplitudes for the Goldstone scalars (or for the zero
helicity, longitudinal gauge bosons) are
$$
a_J(s) = {1\over 32 \pi}\int d(\cos \theta )P_J (\cos \theta )
{\cal{M}}(s,\theta ) \eqno(2.42)
$$
where $\theta$ is the center of mass scattering angle.
Partial wave unitarity then requires
$$
|a_J(s)|\leq 1.\eqno(2.43)
$$
Putting $\rho = 1$ we then find 
$$
|a_0(W_L^+ W_L^-\to Z_LZ_L)| = {s\over 16\pi v^2} \leq 1\eqno(2.44)
$$
so that the interactions of \lag5 must intervene to damp the absolute value of
the amplitude at a scale $\Lambda_{\rm SB}$ bounded by
$$
\Lambda_{\rm SB} \leq 4\sqrt{\pi} v \simeq 1.75 \ TeV.\eqno(2.45)
$$

At the cutoff, $s \cong O(\Lambda_{\rm SB}^{2} )$, the $J=0$ wave is
$$
|a_0 (\Lambda_{\rm SB} )| \cong 
{\Lambda_{\rm SB}^2\over 16\pi v^2}\eqno(2.46)
$$
which relates the strength of the interaction and the energy scale of 
the new physics.  If $\Lambda_{\rm SB} \ \ltap \ {1\over 2} $ TeV then 
$a_0(\Lambda_{\rm SB} )\ \ltap \ 1/4\pi$, well below the unitarity 
limit.  Then $\lag5$ has a weak coupling and can be analyzed 
perturbatively.  For $\Lambda_{\rm SB} \ \gtap \ 1$ TeV, we have 
$|a_0(\Lambda_{\rm SB} )| \ \gtap \ 1/3$, which begins to approach 
saturation.  Then $\lag5$ is a strong interaction theory requiring 
nonperturbative methods of analysis.

The cutoff is accomplished by exchange of quanta in the Higgs boson 
channel (i.e., $J=Q_{\rm EM}=I_{L+R}=0$ where $I_{L+R}$ is the 
`custodial' isospin, the diagonal subgroup of the global $SU2_{L} \times 
SU2_{R}$ discussed in section 2.1.), so we can identify the cutoff 
$\Lambda_{\rm SB}$ with the typical mass scale
$M_{\rm SB}$ of the quanta of $\lag5$,
$$
\Lambda_{\rm SB} \cong M_{\rm SB}.\eqno(2.47)
$$

We can illustrate this with two significant examples.  The first is 
the Weinberg--Salam model, in which $s$--channel Higgs exchange 
provides the cutoff.  Assume that $m_H \gg M_W$ but that $m_H$ is 
small enough that perturbation theory is not too bad --- say $m_H 
\simeq 700$ GeV so that $\lambda /4\pi^2 = m^2_H/8\pi v^2 \simeq 
1/10$.  To leading order the $J=0$ partial wave is
$$
a_0(s) = {s\over 16\pi v^2} - {s\over 16\pi v^2}\  {s\over s-m^2_H}\eqno(2.48)
$$
where the first term arises from ${\cal{L}}_{\hbox{\small{gauge}}}$ 
and the second from the $s$-channel Higgs boson exchange due to 
$\lag5$, now assumed to be the Weinberg-Salam Higgs sector.  For $s\ll 
m^2_H$ the first term dominates, giving the low energy theorem, Eq.  
(2.44), as it must.  But for $s \gg m^2_H$ the two terms combine to 
give 
$$
a_0\bigg\vert_{\displaystyle{s \gg m^2_H}} 
= {m^2_H\over 16\pi v^2}.\eqno(2.49)
$$
Comparing Eq. (2.49) with (2.46) we see that (2.47) is indeed verified, i.e.,
$\Lambda_{\rm SB} \cong m_H$.

Consider next a strongly-coupled example.
In this case we expect to approximately saturate the unitarity bound,
$$
\Lambda_{\rm SB} \cong 4\sqrt{\pi}  v \cong O(2) TeV.\eqno(2.50)
$$
We cannot actually compute $M_{\rm SB}$ in this case but we can relate 
the problem to one that has been studied experimentally.  In hadron 
physics the analogous saturation scale from the $\pi \pi$ low energy 
theorems is
$$
\Lambda_{\hbox{\small{Hadron}}}\cong 4\sqrt{\pi} f_\pi 
\cong 650 MeV\eqno(2.51)
$$
which indeed coincides with the mass scale of the lightest 
(non-Goldstone boson) hadrons, e.g., $m_\rho =$ 770 MeV. The 
coincidence is not surprising: we expect resonances to form when 
scattering amplitudes become strong, as they do at the energy scale of 
the unitarity bound.

The general lesson can be extracted from the Higgs boson example: the 
cutoff occurs at a scale $s \simeq M_{\rm SB}^2$ characteristic of 
\lag5, and at energies $\sqrt{s} \geq M_{\rm SB}$ the magnitude of the 
amplitude is of order
$$
|{a_0(s)}|\cong O\left( {M_{\rm SB}^2 \over (1.8 \hbox{TeV})^2}\right) . \eqno(2.52)
$$
More precisely, $M_{\rm SB}$ is {\em the mass scale of the quanta that 
make the condensate that generates $M_W$ and $M_Z$}.

For $M_{\rm SB} \ll 1.8$ TeV, the Lagrangian \lag5 is weak, $WW$ 
scattering is never strong, and the amplitudes are cut off by the 
exchange of narrow $J=0$ bosons, i.e., {\em Higgs} bosons.  In that 
case \m5 is an appropriately weighted (by vev) average of the Higgs 
boson masses,
$$
M_{\rm SB} = \sqrt{\langle m_H^2 \rangle}. \eqno(2.53)
$$

If $M_{\rm SB} \geq 1$ TeV then \lag5 is strong and Eq.  (2.52) shows 
that there will be strong $WW$ scattering above 1 TeV. We do not then 
necessarily expect a Higgs boson but do expect a complex strongly 
interacting spectrum.  We expect resonances to appear at the mass 
scale $M_{\rm SB}$ at which the partial wave amplitudes become strong, 
$|{a_J(M_{\rm SB}^2)}| \sim O(1)$, which implies $M_{\rm SB} \sim 1\ 
-\ 3$ TeV. As discussed below, strong two body vector resonances up to 
$\simeq 2$ TeV would be observable at the LHC, while if the resonances 
are even heavier we could still probe \lag5 by means of nonresonant 
strong $WW$ scattering.

\noindent {\bf 3. Electroweak radiative corrections and strong  $WW$ 
scattering}

In this section I want to briefly discuss the implications of the 
precision electroweak data.  I will discuss the constraints the 
electroweak data place on the symmetry breaking sector and give an 
estimate of the contribution to $Z$-pole radiative corrections from 
strong $WW$ scattering.  The estimate follows by formulating the 
K-matrix strong scattering model as an effective Higgs boson 
theory.\cite{ehb}.  The result agrees with an estimate by Gaillard 
using the nonlinear sigma model.\cite{mkg} The effective Higgs 
formulation is described in more detail in section 5, where it's 
advantages for the calculation of $WW$ cross sections at high energy 
colliders are discussed.

For several years standard model fits of the precision electroweak 
data have favored a rather light mass for the Higgs boson, of the 
order of 100 - 150 GeV and have strongly excluded the TeV scale.  
Until this summer the strength of that conclusion has been open to 
question.  Two years ago the $R_{b}$ anomaly distorted the global fits 
by favoring a low value for the top quark mass which, because of the 
$m_{t}$ -- $m_{H}$ correlation in the radiative corrections, drove the 
Higgs boson mass to low values.  At the time Dittmaier, Schildknecht, 
and Weiglein\cite{dsw} observed that excluding the $R_{b}$ measurement 
and using the directly measured Fermilab value for $m_{t}$ resulted in 
fits allowing $m_{H}$ to reach the TeV scale.

The quantities in the fit that most directly determine $m_{H}$ are the 
$Z$ boson decay asymmetries.  The left-right polarization asymmetry, 
$A_{LR}$, is the most precise and therefore the most important in the 
fit.  Dittmaier et al.\cite{dsw} and Gurtu\cite{gurtu} both observed 
that in the 1996 data the $A_{LR}$ measurement by itself implied a 
value for $m_{H}$ that conflicted with the lower limit (then 
$m_{H}>65$ GeV) from direct searches at LEP. Dittmaier et al.  found 
that without $A_{LR}$ and $R_{b}$ the fitted value of $m_{H}$ 
increased further, with 900 GeV allowed at the $1\sigma$ CL. Gurtu 
suggested reconciling the conflict by inflating the errors on all the 
aysmmetry measurements and found that the TeV scale was allowed at the 
$2\sigma$ CL.

In the Summer of 1997 the $R_{b}$ anomaly had disappeared, but the 
value of $m_{H}$ did not change much in the global fit, largely due to 
the increased precision of $A_{LR}$.  With improved calculations 
of the radiative ocrrections\cite{imprc} $A_{LR}$ by itself then 
implied a 95\% upper limit on $m_{H}$ at the very
same value (77 GeV) that the direct searches implied a 95\% 
lower limit,\cite{hfits} raising the possibility that the fit was 
skewed to low $m_{H}$.  I constructed a fitting algorithm, best 
formulated in the second of references \cite{hfits}, to incorporate 
the information from the search limits.  The algorithm scales the 
uncertainties of the asymmetry measurements in conflict with the 
search limits by a factor reflecting the aggregate confidence level 
for consistency between the complete set of asymmetry measurements and 
the search limits.  The method is motivated by the $S^{*}$ scale 
factor the PDG\cite{pdg} has long used to fit discrepant data, based 
on their observation that discrepancies occur more often than chance 
expectation and are often with hindsight found to result form 
underestimated systematic errors.  Applied to the 
Summer `97 and Spring `98 data I fits using the algortihm allowed 
(at 95\% CL) values of $m_{H}$ approaching the 
TeV scale, contrary to the conventional global fits.  As you can 
imagine the question of how to carry out the fits in the face of the 
discrepancies with the search limits has been conroversial,  
causing polite disagreement and some bar room brawls.

When the method is applied to the Summer `98 data presented at 
Vancouver\cite{vancouver} the results agree with the conventional 
fits.  From Summer `97 to Summer `98 $A_{LR}$ and 
$A_{FB}^{\tau}$ increased by 1$\sigma$ and 0.5$\sigma$ respectively,  
with half of the shift in $A_{LR}$ occurring in the Spring `98 data. 
As of Summer `98 the measurements still conflict with the search 
limit, each implying $m_{H}<90$ GeV at 88\% CL while the search 
experiments have $m_{H}>90$ GeV at 95\% CL. But the 
aggregate CL for consistency between 
the nine asymmetry measurements and
the search limits increased from $\simeq 0.07$ in Summer `97 to $\simeq 
0.12$ in Spring `98 to $\simeq 0.3$ in Summer `98. The latter CL is 
not small and the algorithm gives essentially no 
correction to the conventional fit.  The algorithm is then 
useful not only for alerting us to possible problems but for 
clarifying when apparent discrepancies are not in fact significant.

The contribution of strong $WW$ scattering to the radiative corrections 
is just one of the ways that a strong symmetry breaking sector 
could affect the electroweak radiative corrections. It is interesting 
to estimate the size of its contribution, though it is important to 
keep in mind that there may be other contributions, possibly with 
different signs, and to take care 
to avoid double counting. Naively one would expect strong $WW$ 
scattering to contribute like a TeV scale Higgs boson. This 
expectation is confirmed by two different estimates, both heuristic, 
which find that the contribution is like that of a 
Higgs boson with mass
$$
m_{H}=\sqrt{{8\sqrt{2}\pi \over G_{F}}} = 4\sqrt{\pi}v 
          \simeq 1.75 {\rm TeV}   
          \eqno{(3.1)}
$$
which is precisely the unitarity cutoff scale defined in eq. 2.45. 
This does not mean that such a Higgs boson would exist but only 
that the contribution to the radiative corrections is like what would 
naively be expected if it made sense to consider a Higgs boson of that 
mass.

This estimate was obtained by Gaillard\cite{mkg} from the nonlinear 
sigma model with a cutoff that is naturally identified with the cutoff 
in eq.  2.45.  I will sketch a heuristic derivation 
using a gauge invariant formulation of strong $WW$ scattering, 
discussed in the next section, in which strong $WW$ scattering models 
are given an effective Higgs boson representation.  Though models of 
strong $WW$ scattering are formulated in renormalizeable gauges using 
the Goldstone boson $ww$ degrees of freedom, it is possible (and 
useful for the study of collider signals) to express the models in 
other gauges including unitary gauge.\cite{ehb}

We consider the K-matrix model, a model of strong $WW$ 
scattering described in section 4, that smoothly extrapolates the 
leading $ww$ partial wave amplitudes from the threshold region, where 
they are given by the low energy theorems, eqs.  2.36 - 2.39, to 
higher energy in a way that exactly satisfies elastic unitarity.  For 
the $I=J=0$ partial wave the K-matrix amplitude is
$$
a_{00}^{K}= {x_{00}\over 1-ix_{00}}   \eqno{(3.2)}
$$
where $x_{00}$ is the low energy theorem amplitude, 
$$
x_{00}={s\over 16\pi v^{2}}.    \eqno{(3.3)}
$$ 
As described in section 5, the corresponding 
effective Higgs propagator is
$$
P^{K}(s)= {i\over s + 16\pi i v^{2}}. \eqno{(3.4)}
$$
Interpreted heuristically as a Breit-Wigner ``resonance'' the pole in 
eq.  3.4 corresponds to a `resonance' whose width is twice its mass, 
$\Gamma_{H}=2m_{H}$.  Used naively to evaluate the one loop radiative 
corrections, the `Higgs' propagator in equation 3.4 induces radiative 
corrections of the standard model form with $m_{H}$ given by eq.  3.1, 
except for a small additional term from the log of the imaginary phase 
of the pole position.

To summarize, it appears even to a skeptic that the precision 
electroweak data now exclude strong symmetry breaking dynamics unless 
associated new physics contributes radiative corrections that offset 
the contribution from strong $WW$ scattering.  It is more natural than 
not that there be additional contributions, and models 
\cite{strongmodels} have been constructed which are consistent with 
the precision data.  We can regard them as existence proofs that 
strong $WW$ dynamics may be consistent with the existing data.  The 
definitive tests require TeV scale high energy colliders, starting 
with the LHC.

\noindent {\bf 4. Models of strong $WW$ scattering}

At the LHC the initial goal of experimental study of $WW$ scattering 
at the TeV scale is to determine whether or not strong scattering 
occurs.  If it does, detailed studies will require even more 
powerful colliders.\cite{snomass} In the spirit of the initial,  
exploratory studies the models of strong $WW$ 
scattering discussed here are not intended as real dynamical theories 
but are meant only to provide estimates of the order of 
magnitude of the expected cross sections in a way that does not 
conflict with general principles such as unitarity.

To get in the spirit of the exercise, the crudest example is the 
linear model\cite{mcmkg2}, that uses the threshold amplitudes, eqs.  
(2.36 - 2.41), which are purely real, as a model of the absolute value 
of the partial wave amplitudes below the unitarity limit and sets the 
absolute value of the partial wave amplitudes equal to one at higher 
energies.  For instance, for the $I,J = 0,0$ partial wave the model is
$$
|{a_{00}}| = {s\over 16\pi v^2} \theta (16\pi v^2 -s) + \theta (s-16 \pi 
v^2).
\eqno(4.1)
$$
The discontinuity in the derivative is unphysical but the model is
nonetheless a potentially useful guide to the {\it magnitude} of certain
partial waves. It gives a surprisingly good description of the
pion scattering data in the $I=J=0$ channel --- see figure 3.2 of 
ref.\cite{wwrev}

K-matrix unitarization is perhaps a step up from the linear 
model,\cite{wwrev_r4} constructed to explicitly satisfy elastic 
unitarity.   Partial wave unitarity is equivalent to
the statement that
$$
Im (a_J^{-1}) = -1 \eqno(4.2)
$$
so that a unitary $a_J$ is completely 
specified by specifying its real part. 
For instance, for the isoscalar channel we choose 
$$
Re (a_{00}^{-1}) = {16\pi v^2 \over s}, \eqno(4.3)
$$
in order to satisfy the low energy theorem. 
The K-matrix amplitude is then 
$$
a_{00} = {s\over 16\pi v^2} \left( 1 - i {s\over 16\pi v^2}\right)^{-1}.
\eqno(4.4)
$$

For the like-charge $I,J = 2,0$ channel the analogous 
model amplitudes are
$$
|{a_{20}}| = {s\over 32\pi v^2} \theta (32\pi v^2 -s) + \theta (s-32 \pi 
v^2)
\eqno(4.5)
$$
and
$$
a_{20} = -{s\over 32\pi v^2} \left( 1 + i {s\over 32\pi v^2}\right)^{-1}.
\eqno(4.6)
$$

Since we are modeling Goldstone boson scattering we can get some 
guidance from the $\pi \pi$ scattering data.  For the $I=1$ and $I=2$ 
amplitudes the data illustrates the complementarity of the resonant 
and nonresonant channels.  In the $I=1$ channel the linear model 
drastically underestimates the magnitude of the amplitude because it 
omits the large enhancing effect of 
the $\rho$ resonance.  In the $I=2$ channel it tracks the data fairly 
well until about $\simeq 600$ MeV (analogous to $\simeq$ 1.6 TeV in 
\wlwl scattering) where it begins to overestimate the data for the 
magnitude of the amplitude.  This is {\em also} a consequence of the 
$\rho$ resonance, which together with the constraints of chiral 
symmetry, {\em suppresses} the $I=2$ amplitude.  If $\rho(770)$ were 
heavier and/or less strongly coupled to $\pi \pi$, the linear model 
would be a better fit to the data in both the isovector and isotensor 
channels.  The $a_{11}$ amplitude would then be smaller while 
$|a_{20}|$ would be bigger!

The chiral lagrangian with the \ro meson incorporated in a chiral 
invariant fashion is a useful tool to illustrate complementarity of 
resonant and nonresonant scattering signals.  
Weinberg\cite{wwrev_25} showed that chiral invariance requires 
the conventional \rpp interaction,
$$
{1\over 2}f_{\rho \pi\pi}\ \epsilon_{ijk}\ \rho^{\mu}_i \pi_j
{\stackrel{\leftrightarrow}\partial}_{\mu} \pi_k,\eqno(4.7)
$$
\begin{figure}
\epsfbox{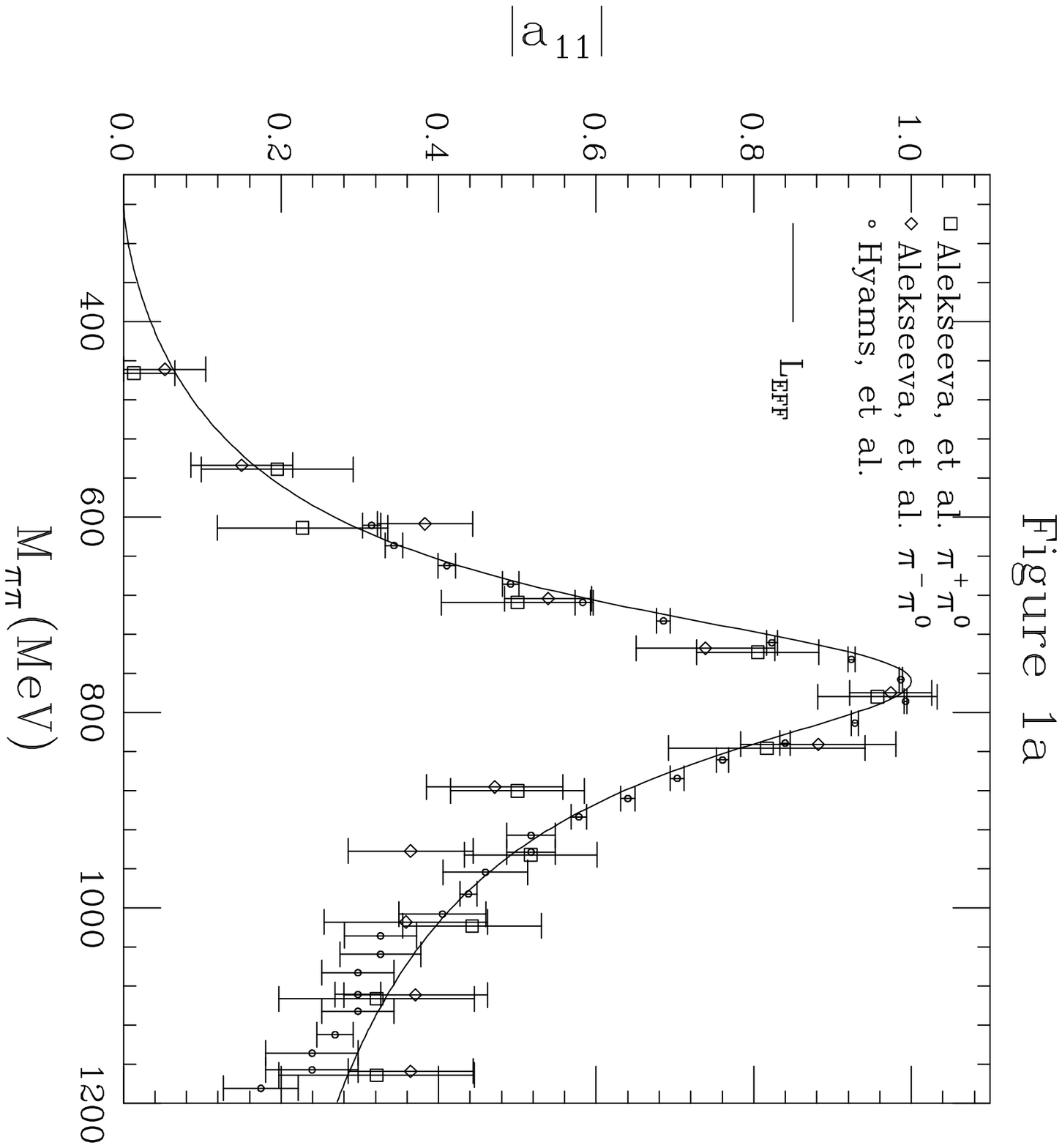}
\end{figure}
\begin{figure}
\epsfbox{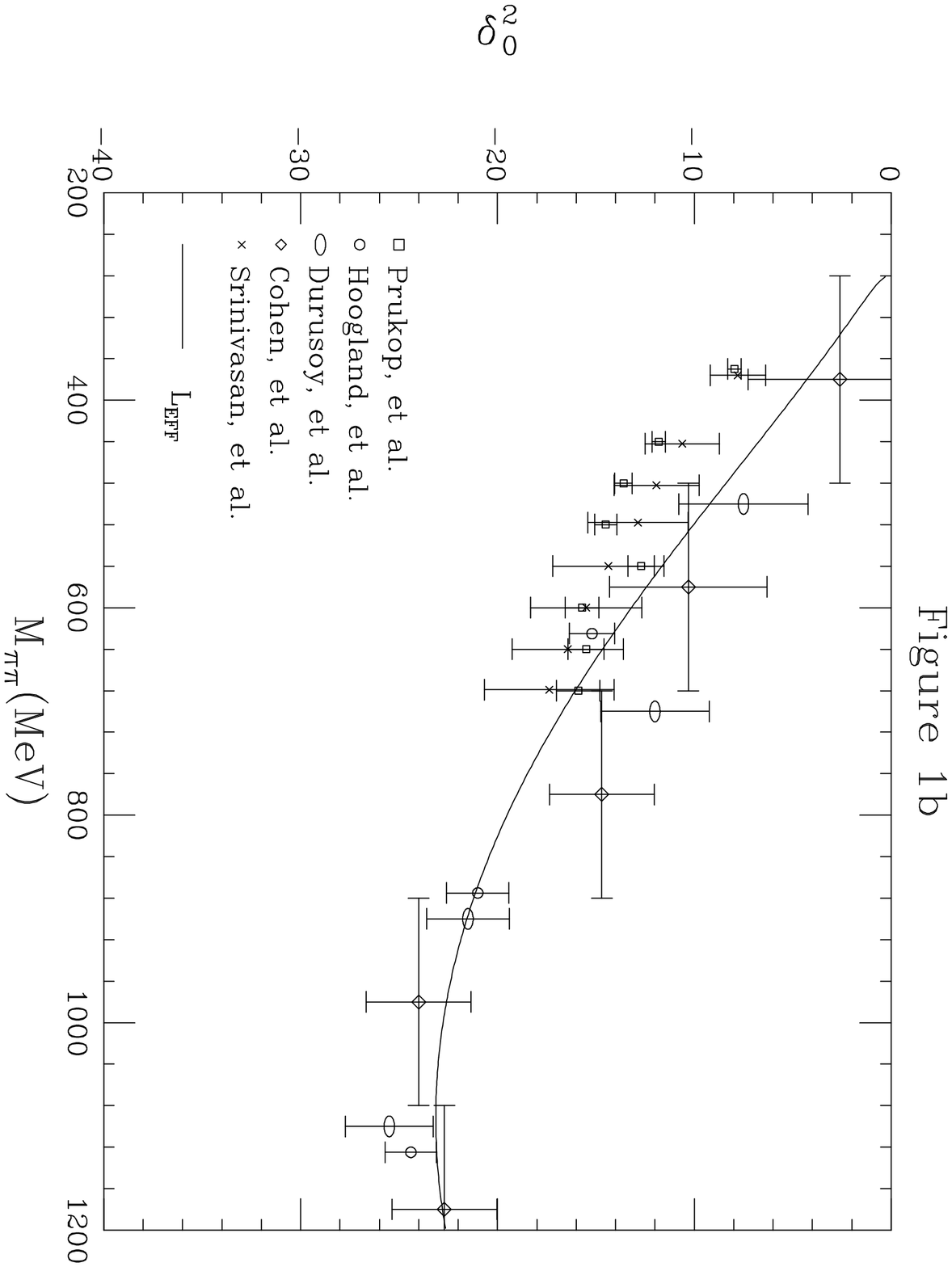}
\end{figure}
\noindent 
to be accompanied by a four pion contact interaction that induces a 
term linear in $s$ in the \pp scattering amplitude.  This term cancels 
the linear terms induced by \ro exchanges, so that the low energy 
theorems are guaranteed.

This chiral lagrangian gives a remarkably good fit (with no free 
parameters) to the \pp data for the $a_{11}$ and $a_{20}$ partial 
waves, for energy as large as 1.2 GeV, shown in figure 1.\cite{mcwk1}
The quality of the fit at 1.2 GeV must be fortuitous 
since the linear terms contributed by the chiral lagrangian are 
irrelevant at that scale.  Nevertheless the chiral lagrangian gives 
a good parameterization of the QCD data, 
which we can use to explore the 
consequences of varying the \ro mass and width.  For massless 
Goldstone bosons $w_i$ the \ro width is given by
$$
\Gamma_{\rho}={f_{\rho ww}^2\over 48\pi}m_{\rho}.\eqno(4.8)
$$
The Lagrangian is completely specified by the Goldstone boson -- gauge 
current coupling $v$, and the \ro mass and width.  (With an additional 
direct \ro coupling to fermions, it is also the basis of the BESS 
model.\cite{bess})

To represent possible $WW$ vector resonances we consider two examples 
of \qrho mesons from minimal, one doublet technicolor, $N_{TC}=2$ and 
4.  With the conventional large $N_{TC}$ scaling the mass and width 
are given in terms of the parameters of the $\rho (770)$ by
$$
m_{\rho_{TC}}=\sqrt{{3\over N_{TC}}}{v\over F_{\pi}}m_{\rho}
\eqno(4.9)
$$ 
and 
$$
\Gamma_{\rho_{TC}}={3\over N_{TC}}{m_{\rho_{TC}}\over m_{\rho}}
{1\over \beta_{\pi}^3}\Gamma_{\rho}.\eqno(4.10)
$$
where $\beta_{\pi}$ is the pion velocity in the $\rho (770)$ decay.  
For $N_{TC}=4$ the mass and width are 1.78 and 0.326 TeV. For 
$N_{TC}=2$ they are 2.52 and 0.92 TeV. For larger values of $N_{TC}$ 
and with the addition of more techniquark doublets the \rhotc becomes 
lighter and more easily observable.  To represent the possibility that 
the resonances of \lag5 may be heavier than the naively anticipated 1 
- 3 TeV region I also consider a 4 TeV \qrho meson with a width of 
0.98 TeV determined from the $f_{\rho \pi \pi}$ coupling of the $\rho 
(770)$.

The amplitudes are unitarized by the K-matrix method.  The width is 
omitted in the real part of the s-channel pole contribution and the 
imaginary part (i.e., the width) is determined from the K-matrix 
prescription.  This is equivalent to the conventional broad resonance 
Breit-Wigner parameterization with the fixed imaginary part of the 
denominator, $m \Gamma$, replaced by $\sqrt{s} \Gamma(\sqrt{s})$.

The complementarity of the $a_{11}$ and $a_{20}$ channels is evident 
in figure 2.\cite{mcwk1} As $m_{\rho}$ increases, the \ro 
Lagrangian amplitudes approach the nonresonant K-matrix model 
amplitude for $|a_{11}|$ from above and $|a_{20}|$ from below, since 
chiral invariant 
\ro exchange enhances the former and suppresses the latter.  At the 
LHC the 4 Tev \qrho signal is indistinguishable from the signal of  the 
nonresonant K-matrix model.  The fact that the \qrho resonance 
amplitude approaches the nonresonant K-matrix amplitude for large \qrho 
mass is a very general feature, independent of the specific properties 
of vector meson exchange.  It explains the sense in which smooth 
unitarization models, such as the linear and K-matrix models, are 
conservative: 
\begin{figure}
\epsfbox{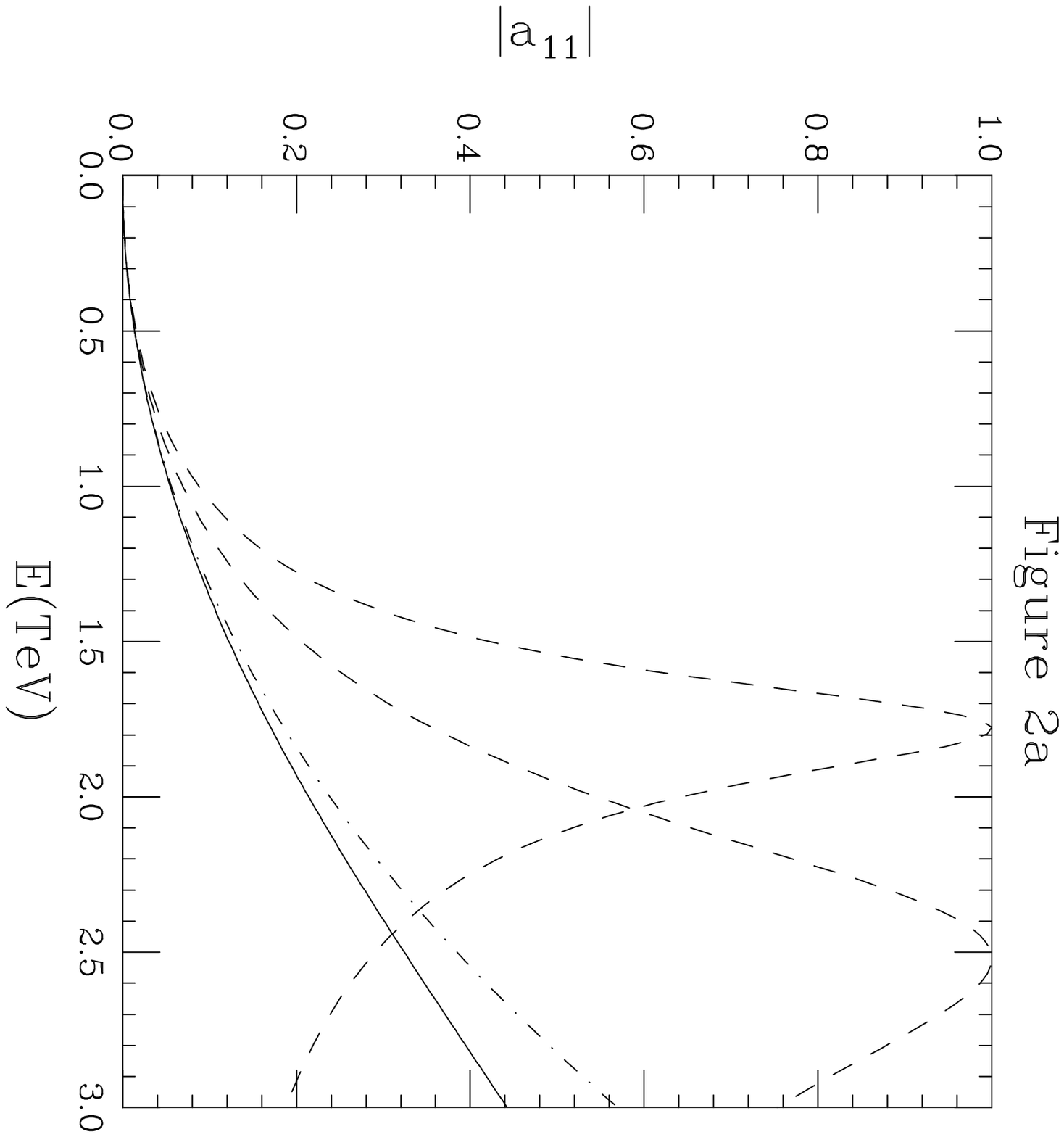}
\end{figure}
\begin{figure}
\epsfbox{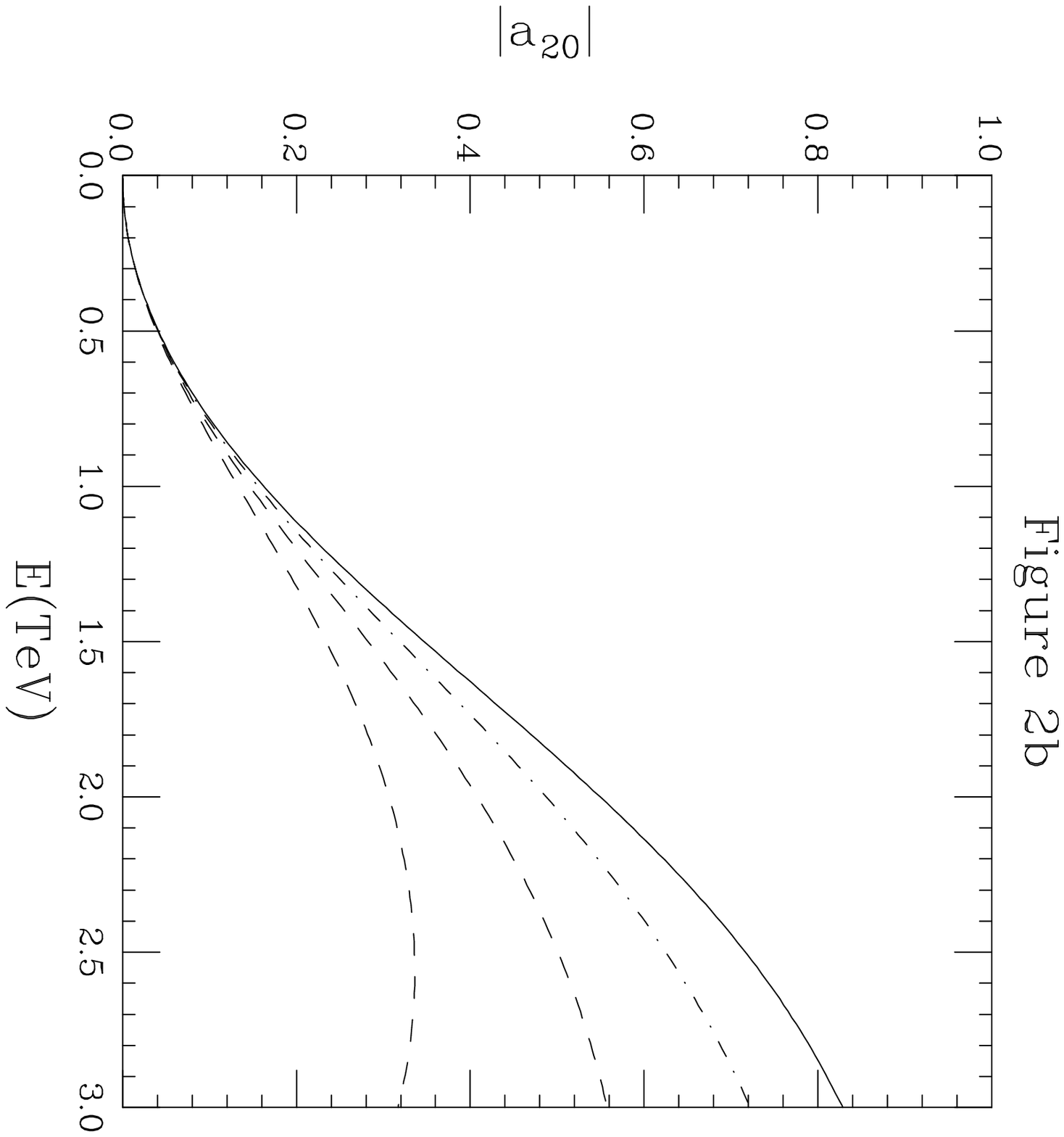}
\end{figure}
\noindent 
they represent the ``fail-safe'' nonresonant 
scattering signals that are anticipated if the resonances are 
unexpectedly heavy.  This is the most general meaning of 
complementarity.  A more specific meaning, special to vector meson 
exchange as constrained by chiral symmetry, is the inverse 
relationship of the $I=1$ and $I=2$ channels.

\noindent {\bf 5. Strong $WW$ scattering cross sections at high energy 
colliders: EWA and EHB}

This section is concerned with tools to translate models of $WW$ 
scattering into collider cross sections.  The now traditional method 
is the effective $W$ approximation\cite{ewa} (EWA), which is 
computationally convenient and sufficiently accurate within the 
experimentally relevant domain of applicability.  However, 
because it is a small angle approximation the EWA provides no 
information on the experimentally important transverse momentum 
spectrum of the final state jets or of the diboson in the $qq \ra 
qqWW$ process used to observe $WW$ scattering at hadron colliders.  In 
addition to reviewing the EWA I will also describe the effective 
Higgs boson method\cite{ehb}, which provides complete information on 
the final state particles and also retains interference of signal and 
background amplitudes that is neglected by the EWA. 

\noindent {\em 5.1 Effective $W$ approximation}

The effective $W$ approximation is analogous to the effective photon 
approximation of Weiszacker and Williams.  It provides an effective 
luminosity distribution for the probability to find colliding 
``beams'' of longitudinally polarized gauge bosons within the 
colliding quark ``beams'' produced at a $pp$ collider or within the 
$l^{+}l^{-}$ beams at a lepton collider.  For incident fermions $f_1$ 
and $f_2$ the effective luminosity for longitudinally polarized gauge 
bosons $V_1$ and $V_2$ is
$$
{\partial{\cal{L}}\over \partial z}\Bigg|_{V_1V_2/f_1f_2} = 
{\alpha^2\chi_1\chi_2 \over \pi^2 \sin^4\theta_W}{1\over z} \left[(1+z) 
\ln \left({1\over z}\right) - 2+2z\right]\eqno(5.1)
$$
where $z\equiv s_{VV}/s_{ff}$ and the $\chi_i$ are the $f_i$ - $V_i$ 
couplings, {e.g.}, $\chi_W = 1/4$ for all fermions, $\chi_{Zu\bar{u}}= 
(1+(1-{8\over 3}\sin^2\theta_W)^2)/16 \cos^2\theta_W$, etc $\ldots$.  
Equation (5.1) must be convoluted with the desired $V_1V_2$ subprocess 
cross section and also with the quark distribution functions in the 
case of $pp$ collisions,
$$
\sigma (pp\to \cdots)
=\int_\tau\left.{\partial{\cal{L}}\over \partial 
\tau}
\right|_{qq/pp}\cdot\int_x
\sum_{V_L}
\left.{\partial{\cal{L}}\over \partial 
x}\right|_{V_LV_L/qq}\cdot\sigma (V_LV_L\to \cdots)\eqno(5.2)
$$
The effective $W$ approximation has been compared with analytical and 
numerical evaluations of Higgs boson production.  The analytical 
calculations\cite{wwrev_24} show good agreement for $WW\to H$ for $m_H 
\geq 500$ GeV, with errors $\ltap \;O(10\%)$ and decreasing with $m_H$ 
and $\sqrt{s}$, while for the relatively less important process $ZZ\to 
H$ the errors are roughly twice as large.  Above 1 TeV the errors are 
very small.

\noindent {\em 5.2 Effective Higgs boson}

Because it is a small angle approximation the EWA tells us nothing 
about the transverse momentum of the final state quark jets.  As 
discussed in section 6, a central jet veto and/or forward jet tag are 
useful experimental strategies, requiring knowledge of both signal and 
background jet $p_{T}$ distributions.  Common practice is to use the 
EWA while assuming that the jet $p_{T}$ distribution for strong $WW$ 
scattering is the same as that of the standard model with a 1 TeV 
Higgs boson.  Near the edge of phase space, as we are in the study of 
the $WW$ system at $\geq 1$ TeV at the LHC, the the jet $p_{T}$ 
distribution varies with the $WW$ invariant mass.  The EHB 
method\cite{ehb} provides the appropriate final state $p_{T}$ 
distribution for any s-wave strong scattering model.  In 
addition, because the collider signal is computed from the model {\em 
amplitude}, not the cross section as in eq.  5.2, the EHB retains 
phase information and signal-background interference effects that are 
lost in the EWA.

Consider a strong s-wave scattering model `$X$', formulated in an 
R-gauge as a Goldstone boson scattering amplitude, for instance, the 
leading $J=0$ component of $\caM^{X} (w^{+}w^{-} \ra zz)$.  To leading 
order in the $SU(2)_L$ coupling constant $g$ we decompose the 
corresponding $W_{L}^{+}W_{L}^{-}\ra Z_{L}Z_{L}$ amplitude into gauge 
sector and symmetry-breaking sector contributions,
$$
{\cal M}^{X}_{\rm Total}(W_{L}W_{L}) = 
{\cal M}_{\rm Gauge}(W_{L}W_{L})
+ {\cal M}^{X}_{\rm SB}(W_{L}W_{L}).
\eqno(5.3)
$$
${\cal M}_{\rm Gauge}$ is the sum of $W,\gamma$, and $Z$ 
exchange diagrams; it increases like $E^{2}$ for large $E$, 
$$
{\cal M}_{\rm Gauge}= g^2{E^2 \over M_W^2} 
+ {\rm O}(g^{2},E^{0}).
\eqno(5.4)
$$

This order $E^{2}$ term is precisely the ``bad high energy behavior'' 
discussed in section 2 that 
${\cal M}^{X}_{\rm SB}(W_{L}W_{L})$ must cancel.  It 
is also precisely the low energy theorem amplitude, as can be seen by 
comparing eq.  5.4 with eqs.  2.34 and 2.36.  This is no 
coincidence\cite{let}: if the symmetry breaking force is strong, the 
quanta of the symmetry breaking sector  are heavy, $M_{SB} \gg M_W$, 
and decouple in gauge boson scattering at low energy, ${\cal 
M}^{X}_{SB} \ll {\cal M}_{\rm Gauge}$.  Then the quadratic term in 
${\cal M}_{\rm Gauge}$ dominates ${\cal M}_{\rm Total}$ for $M_W^2 
\ll E^2 \ll M_{SB}^2$, which establishes the low energy theorem to 
order $g^2$ without using the ET. Thus we may also write
$$
{\cal M}_{\rm LET} = {s \over v^2} = 
{\cal M}_{\rm Gauge} + {\rm O}(g^2,E^{0}).
\eqno(5.5)
$$

We next use the equivalence theorem to assert the approximate equality 
of the U-gauge amplitude with the model amplitude formulated in R-gauge,
$$
{\cal M}^{X}_{\rm Total}(W_{L}W_{L}) = 
{\cal M}^{\rm X}_{\rm Goldstone}(ww)  + {\rm O}(g^{2},E^{0}).
\eqno(5.6)
$$
Combining eqs.  5.4 - 5.6 we find the corresponding contribution of 
the symmetry breaking sector to the U-gauge 
$W_{L}W_{L}$ amplitude,
$$
{\cal M}_{\rm SB}^{\rm X}(W_LW_L) = {\cal M}^{\rm X}_{\rm Goldstone}(ww) - 
{\cal M}_{\rm LET} + {\rm O}(g^2,{M_W \over E}).
\eqno(5.7)
$$

Finally we reexpress eq 5.7 in terms of an effective Higgs boson 
propagator, $P^{X}_{\rm EFF}$, constructed to be exchanged in the 
s-channel with standard model $WW$ and $ZZ$ couplings,
$$
P^{X}_{\rm EFF}(s) = -i {v^2 \over s^2}
                  ({\cal M}^{\rm X}_{\rm Goldstone}(ww)
                 - {\cal M}_{\rm LET}),
\eqno(5.8)
$$
defined so that its $s$-channel exchange reproduces equation (5.7).  
Notice that ${\cal M}_{\rm LET}$ contributes $i/s$ to $P_{\rm EFF}^X$, 
corresponding to a massless scalar pole, making explicit the 
connection between the spontaneously broken symmetry that implies 
${\cal M}_{\rm LET}$ and the cancellation of the bad high energy 
behavior by Higgs boson exchange.  The residual contribution to 
$P_{\rm EFF}^X$ from ${\cal M}^X_{\rm Goldstone}$ carries the model 
dependent strong interaction dynamics.

We can use $P^{X}_{\rm EFF}(s)$ in any gauge to extract the 
consequences of model $X$.  To compute the predicted collider signal, 
we compute the full standard model set of Feynman diagrams for $qq \ra 
qqWW/ZZ$ using $P^{X}_{\rm EFF}(s)$ for the Higgs boson propagator.  
It is shown in references \cite{ehb} that the method agrees with the 
EWA where it should, while improving on the EWA at small angles where 
it includes interference with the photon t-channel exchange amplitude 
that is neglected in the EWA. Just as in the tree-level evaluation of 
SM Higgs boson production, it provides a complete description of the 
final state.  It is also shown, in the second of references\cite{ehb}, 
that the amplitudes are BRS invariant.

\noindent {\bf 6. Can LHC lose?}

This section offers a sketch of strong $WW$ scattering at the LHC, 
with the strategies to enhance the signals relative to the backgrounds 
and an estimate of the integrated luminosity needed to either confirm 
or exclude the signals.  The discussion will focus on the chiral \qrho 
effective Lagrangian.  As described in section 4 it illustrates the 
complementary interplay between resonant and nonresonant signals, and 
in the $m_{\rho}\ra \infty$ limit approaches the nonresonant K-matrix 
model.  We consider $WZ$ scattering which the \qrho resonance would 
enhance and like-charge $W^{+}W^{+}$ scattering which chiral invariant 
\qrho exchange would suppress.  Discussion of other models and 
channels may be found in references \cite{jbetal,mc_otherww}.

\noindent {\em 6.1 $W^{+}W^{+}$ elastic scattering}

The $W^+W^+$ and $W^-W^-$ channels are 
interesting for three reasons:
\begin{itemize}
\item They do not have the \qqb $\to WW$ or $gg\to WW$ backgrounds, 
respectively of order $\alpha_W$ and \alphawalphas in amplitude, that 
are the dominant backgrounds to strong scattering in other 
gauge boson pair channels.  
 \item The branching ratio for $W^+W^+\to l^+\nu + l^+\nu$ with $l = e$
or $\mu$ is relatively large, $\sim $ 5\%, and has a striking experimental
signature: two isolated, high $p_T$, like-sign leptons in an event 
with no other significant activity (jet or leptonic) in the central 
region.

\item Strong $W^+W^+ + W^-W^-$ scattering complements the strong 
scattering signals in the other gauge boson pair channels, and is 
likely to be largest if the resonance signals expected in the other 
channels are smallest.

\end{itemize}

The \wpwp strong scattering signal was first estimated in 
ref.\cite{mcmkg2} but with no estimate of the backgrounds.  There 
have subsequently been several more detailed studies of signals and 
backgrounds,\cite{wwrev_456} resulting in a powerful set of 
cuts.  One would expect the $O(\alpha_S\alpha_W)$ gluon exchange 
amplitude for $qq \to qqW^+W^+$ to be the dominant background, but, 
surprisingly, after cuts it is much smaller than the electroweak 
$O(\alpha^2_W)$ background.\cite{wwrev_37} Even more surprising, 
$\overline qq\ra W^{+}Z$\cite{azuelosetal,mcwk2} and $\overline qq\ra 
W^{+}\gamma^{*}$\cite{mcwk2} with the negative charge lepton escaping 
detection are as important as the irreducible $qq \to qqW^+W^+$ 
backgrounds.  (Backgrounds from $W^+W^-$ production with 
mismeasurement of a lepton charge and top quark related 
backgrounds\cite{wwrev_r4,ttb} can be controlled and 
are not considered here.)

While a forward jet tag may provide further background suppression, 
the results quoted below, from \cite{mcwk2}, rely only on hard lepton 
cuts and a central jet veto (CJV) of events containing a jet with 
central rapidity, $\eta_{J}<2.5$ and high transverse momentum, 
$P_{T}(J) > 60$ GeV.  The CJV reduces backgrounds from 
transversely polarized $W$ bosons, which are emitted at larger 
transverse momenta than the longitudinally polarized $W$ bosons of the 
signal.  The hard lepton cuts rely on the general property that the 
strong scattering subprocess cross sections increase with $s_{WW}$ 
while the backgrounds scale like 1/$s_{WW}$, and on the differing 
polarization of the signal and background $WW$ pairs.  If this strategy 
suffices it has the advantage of being cleaner than relying 
on forward jet tagging, which may be subject to QCD 
corrections and to detector-specific jet algorithms and acceptances in 
the forward region.

The leptonic cuts are optimized for each set of model parameters.  It 
turns out for the four values of $m_{\rho}$ considered that the 
rapidity and lepton transverse momentum cuts are $\eta( l)<1.5$ 
and $p_{T}(l) >130$ GeV. A third cut, requiring the two leptons 
to be back-to-back in azimuth, depends somewhat on $m_{\rho}$.

The $WZ$ and $W\gamma^{*}$ backgrounds --- actually the complete 
background from all amplitudes for 
$\overline qq \rightarrow l^+ \nu l^+ l^-$ --- in 
which the $l^-$ escapes detection, occurs because any detector has 
unavoidable blind spots at low transverse momentum and at high 
rapidity.  At very low $p_T$, muons will not penetrate the muon 
detector, electrons or muons may be lost in minimum bias pile-up, and 
for low enough $p_T$ in a solenoidal detector they will curl up 
unobservably within the beam pipe.  Muon and electron coverage is also 
not likely to extend to the extreme forward, high rapidity region.

In reference \cite{mcwk2} an attempt was made to employ reasonable 
though aggressive assumptions about the observability of the extra 
electron or muon.  Rapidity coverage for electrons and muons was 
assumed for $\eta(l) < 3$.  Within this rapidity range it was assumed 
that isolated $e^-$ and $\mu^-$ leptons with $p_T(l) > 5$ GeV can be 
identified in events containing two isolated, central, high $p_T$ 
$e^+$'s and/or $\mu^+$'s.  It was also assumed that electrons (but not 
muons) with $1< p_T(l) < 5$ GeV can be identified if they are 
sufficiently collinear ($m(e^+e^-) < 1$ GeV) with a hard positron in 
the central region.  For $p_T(e^-) < 1$ GeV electrons were considered 
to be unobservable. 

A robust observability criterion is defined and the cuts are optimized 
by searching over 
the cut parameter space for the set of cuts that 
satisfy the observability criterion with the smallest integrated 
luminosity.  The criterion is 
$$
\sigma^{\uparrow}   =  S/\sqrt{B}  \ge  5 
\eqno(6.1)
$$
$$
\sigma^{\downarrow}   =  S/\sqrt{S+B}  \ge  3
\eqno(6.2)
$$
$$
S \ge B,
\eqno(6.3)
$$
where $S$ and $B$ are the number of signal and background events, and 
$\sigma^{\uparrow}$ and $\sigma^{\downarrow}$ are respectively the 
number of standard deviations for the background to fluctuate up to 
give a false signal or for the signal plus background to fluctuate 
down to the level of the background alone.  The $\sigma^{\downarrow}$ 
criterion is essential to assure the ability to exclude strong 
scattering if it does not exist.  In addition 
$S \ge B$ is required so that the signal is unambiguous despite the 
systematic uncertainty in the size of the backgrounds, which will 
probably be known to better than $ \leq \pm 30 \%$ after 
``calibration'' studies
with known processes at the LHC. An experiment 
meeting this criterion, eqs. 6.1 - 6.3, can 
defintitively establish the existence 
\newpage 
\begin{small}
\begin{quotation}
\noindent {\bf Table 6.1} Minimum integrated luminosity 
${\cal{L}}_{MIN}$ to satisfy significance criterion for $W^+W^+ + 
W^-W^-$ scattering.  Also shown are the optimum cuts, the 
corresponding number of signal and background events per 100 
fb$^{-1}$,  
and the compositon of the background for the optimum cut.  Rejection 
of all events for which the third (wrong-sign) lepton falls within its acceptance 
region is assumed. 
\end{quotation}
\end{small}
\begin{center}
\begin{tabular}{ccccc}
$m_{\rho}$(TeV) & 1.78 & 2.06&2.52&4.0\cr
\hline
\hline
${\cal{L}}_{MIN}\ ({\rm{fb}}^{-1})$&142&123&105&77\cr
\hline
$WW$ Cut&&&&\cr
$\eta^{MAX}(l)$&1.5&1.5&1.5&2.0\cr
$p^{MIN}_{T}(l)$ (GeV)&130.&130.&130.&130.\cr
$[\cos\phi(ll)]^{MAX}$&$-0.72$&$-0.80$&$-0.80$&$-0.90$\cr
\hline
$WW$ Sig/Bkgd&&&&\cr
(events per $100\ {\rm{fb}}^{-1})$&12.7/6.0&14.1/5.8&15.9/5.8&22.4/8.9\cr
\hline
$WW$ Backgrounds (\%)&&&&\cr
$\overline l l \overline l \nu_{l}$&47&49&49&61\cr
$O(\alpha^2_W)$&47&46&46&33\cr
$O(\alpha_W\alpha_S)$&6&6&6&6\cr
\hline
\hline
\end{tabular}
\end{center}
\noindent 
of  strong scattering if it exists or exclude it if it does not. As 
discussed in the introduction, the latter capability is as important 
as the former. The criteria are applied to the actual event 
yields after correcting  
for detector efficiency, assumed to be 85\% for each 
isolated lepton falling within the region defined by the cuts. 

The results are collected in table 6.1 for four values of the \qrho 
mass.  We see that the heaviest value of $m_{\rho}$ gives the largest 
signal, requiring the smallest integrated luminosity, ${\cal L}_{MIN}= 
77$ fb$^{-1}$, less than a year of running at the design luminiosty of 
$10^{34}$ cm$^{-2}$sec$^{-1}$.  The nonresonant K-matrix and linear 
models provide similar though slightly bigger signals.  For the 
lightest mass considered, about 1${1\over 2}$ years would be needed.  
In tables 6.1 - 6.3 the event yields per 100 fb$^{-1)}$ do not include 
detector efficiency while ${\cal L}_{MIN}$ does.

Table 6.1 assumes 100\% veto efficiency when the
third lepton in the $l^+ \nu l^+ l^-$ background falls within the 
geometric acceptance specified above.  Table 6.2 shows the effect of 
veto inefficiency for $m_{\rho}=2.52$ TeV. At 98\% efficiency the 
effect is not great but at 95\% ${\cal L}_{MIN}(WZ)$ is increased by
40\%.  For 90\%, not shown in the table, ${\cal L}_{MIN}(WZ)$ would be 
nearly doubled, to 200 fb$^{-1}$. Hopefully  an aggressive 
$\simeq 98\%$ efficient veto is possible without significantly affecting 
the signal efficiency. The question clearly depends on the 
capability of the particular detectors. 

\noindent {\em 6.2 The $WZ$ signal}

The $WZ$ signal arises from two separate mechanisms.\cite{mcmkg2} The 
first is elastic $WZ$ scattering via $qq\to qqWZ$ where the $WZ\to WZ$ 
subprocess is mediated by s-channel and u-channel $\rho$ exchange as 
well as the contact interactions required by chiral 
symmetry.\cite{mcwk1} The second is by $\overline qq \to \rho$, 
evaluated using $\rho$ dominance.  The backgrounds are $\overline qq 
\rightarrow WZ$ and $qq \to qqWZ$, the latter from both the 
$\hbox{O}(\alpha_W^2)$ and $\hbox{O}(\alpha_W\alpha_{S})$ amplitudes.  
The $\hbox{O}(\alpha_W^2)$ background 
amplitude is the $qq \to qqWZ$ cross 
section from $SU(2) \times U(1)$ gauge interactions, computed in the 
standard model with a light Higgs boson, say $m_H \le 0.1$ TeV.

\newpage
\begin{small}
\begin{quotation}
\noindent {\bf Table 6.2} Minimum luminosity to satisfy significance 
criterion for $W^+W^+ + W^-W^-$ scattering for $m_{\rho}=2.52$ TeV, 
assuming 100\%, 98\% or 95\% efficiency for the
veto of wrong-sign charged leptons that fall
within the acceptance region specified in
the text. The optimum cuts and corresponding yields are 
shown as in table 1.
\end{quotation}
\end{small}
\begin{center}
\begin{tabular}{ccccc}
Efficiency & 100\% & 98\% & 95\% \cr
\hline
\hline
${\cal{L}}_{MIN}(WW)\ ({\rm{fb}}^{-1})$&105&115&148 \cr
\hline
$WW$ Cut&&&&\cr
$\eta^{MAX}(l)$&1.5&1.5&1.5\cr
$p^{MIN}_{T}(l)$ (GeV)&130.&130.&160.\cr
$[\cos\phi(ll)]^{MAX}$&$-0.80$&$-0.80$&$-0.86$\cr
\hline
$WW$ Sig/Bkgd&&&&\cr
(events/$100\ {\rm{fb}}^{-1})$&15.9/5.8&15.9/7.9&12.1/5.6\cr
\hline
$WW$ Backgrounds (\%)&&&&\cr
$\overline l l \overline l \nu_{l}$&49&62&71\cr
$O(\alpha^2_W)$&46&34&26\cr
$O(\alpha_W\alpha_S)$&6&4&3\cr
\hline
\hline
\end{tabular}
\end{center}

The results presented here are taken from reference \cite{mcwk1}.  The 
signal is observed in the decay channel $WZ \rightarrow l\nu + 
\overline ll$ where $l = e$ or $\mu$, with net branching ratio $BR = 
0.0143$.  A central jet veto is applied as in section 6.1.  Because 
the signals occur at enormous $WZ$ energy they stand out 
prominently and simple cuts suffice --- on the lepton 
rapidity $\eta_{l}< \eta_{l}^{\rm MAX}$, the azimuthal angles 
$\phi_{ll}$ between the leptons from the $Z$ and the charged lepton 
from the $W$, $\hbox{cos}\phi_{ll} < (\hbox{cos}\phi_{ll})^{MAX}$, and 
the $Z$ transverse momentum, $p_{TZ} > p_{TZ}^{MIN}$.  The cuts are 
optimized for each choice of $m_{\rho}$ to minimize the integrated 
luminosity satisfying eqs.  6.1 - 6.3.  The detector efficiency for 
$WZ \rightarrow l\nu + \overline ll$ is estimated\cite{mcwk1_19} to be 
$0.85 \times 0.95 \simeq 0.8$.

The results are summarized in table 6.3, for three values of 
$m_{\rho}$.  The signal for $m_{\rho}=1.78$ TeV is easily visible, 
meeting the observability criterion with less than a half year at 
design 

\begin{small}
\begin{quotation}
{\bf Table 6.3} Minimum luminosity to satisfy observability criterion 
for $W^{\pm}Z$ scattering for $m_{\rho}=1.78, 
2,52,4.0$ TeV. Each entry displays ${\cal L}_{MIN}$ in fb$^{-1}$, the 
number of signal/background events per 100 fb$^{-1}$, and the 
corresponding values of the cut parameters $\eta^{MAX}(l)$, 
$p_{TZ}^{MIN}$, and ${\rm cos}(\phi_{ll})^{\rm MAX}$.  A central jet 
veto is applied as discussed in the text. 
\end{quotation}
\end{small}
\begin{center}
\begin{tabular}{cccc}
    $m_{\rho}$(TeV) &1.78 TeV & 2.52 TeV & 4.0 TeV \\
\hline
\hline    
     ${\cal{L}}_{MIN}(WZ)\ ({\rm{fb}}^{-1})$&44 fb$^{-1}$  
                  & 300 fb$^{-1}$ & no signal \\
\hline
Cuts &&&\\
$\eta^{MAX}(l)$&2&2&\cr
$p^{MIN}_{TZ}(l)$ (GeV)&450&675&\cr
$[\cos\phi(ll)]^{MAX}$&$1.0$&$1.0$&\cr
\hline 
$W^{\pm}Z$ Sig/Bkgd &&&\\
(events/100 fb$^{-1}$)& 38/20 & 5.8/3.4 & \\
\hline
\hline 
\end{tabular}
\end{center}

\newpage

\noindent luminosity.  (The signal for $m_{\rho}=2.06$ TeV 
[corresponding to $SU(3)_{TC}$] is not shown in the table; it requires 
98fb$^{-1}$ to meet the criterion.)  The heaviest mass considered, 
$m_{\rho}=4$ TeV, is indistinguishable at the LHC from nonresonant 
scattering models; it provides no signal consistent with eq.  6.3.  
The $m_{\rho}=2.52$ TeV signal ($SU(2)_{TC}$) requires three years at 
design luminosity.

\noindent {\em 6.3 The bottom line}

Comparing tables 6.1 and 6.3 the complementarity of the nonresonant 
$W^{+}W^{+}$ channel and the resonant $WZ$ channel is clear.  The 
lightest value of $m_{\rho}$ provides the smallest $W^{+}W^{+}$ signal 
and the largest, readily observable $WZ$ signal.  Models with very heavy 
$m_{\rho}$ or of nonresonant strong scattering cannot be observed in 
the $WZ$ channel for any luminosity while satisfying eq.  6.3, but they 
provide the largest signals in $W^{+}W^{+}$ scattering.  

The over-all worst case, intermediate between these extremes, 
is the value of $m_{\rho}$ for which 
MIN(${\cal{L}}_{MIN}(WZ), {\cal{L}}_{MIN}(W^{+}W^{+})$) is largest.  
This turns out to be $m_{\rho}=2.52$ TeV, corresponding to 
$SU(2)_{TC}$, for which the best signal is in the $W^{+}W^{+}$ 
channel.  It determines the ``no-lose'' luminosity needed to observe 
strong scattering in at least one channel.  As shown in table 6.2, 
this luminosity depends on the experimental veto 
efficiency for wrong-sign leptons that fall within the experimental 
acceptance.  Following the slogan ``when in doubt throw it out,'' it 
may be possible to achieve 98\%
veto efficiency without significantly eroding the efficiency for the 
signal, in which case little more than one year at design luminosity 
would suffice.  More pessimistically, with 95\% veto efficiency, one 
and a half years would be needed.

\noindent {\bf 7. Conclusion}

Only direct discovery and detailed study of the symmetry breaking 
quanta can establish the nature of the symmetry breaking sector in a 
model-indpendent way.  The ability to measure strong $WW$ scattering 
is an important part of the experimental program whether electroweak 
symmetry breaking is strong or not.  Given this capability we can 
determine the mass scale of the symmetry breaking sector even if its 
constituent quanta initially escape detection at the LHC. And even if 
a light Higgs boson is discovered, we would want to verify the absence 
of strong $WW$ scattering, a fundamental prediction of theories in 
which symmetry breaking is dominated by light Higgs bosons. 

The strong $WW$ scattering signals are challenging: they push the LHC 
to the limits of its reach.  The theoretical estimates of the 
possiblity of detecting strong $WW$ scattering at the LHC may seem 
simple and optimistic. But, as we have learned from 
the Higgs boson searches at LEP, experimenters working 
with real detectors and real data can invent and validate clever 
strategies that exceed even the most optimistic of the early 
theoretical simulations.

Can LHC lose?  --- No, not likely, as long as the accelerator and 
detectors succeed in reaching  the ambitious design goals.

\vskip .2in
\noindent{\bf Acknowledgements:} I wish to thank Dirk Graudenz, Milan 
Locher, and Christine Kunz for a very stimulating and pleasant week.
This work was supported by the Director, Office of Energy 
Research, Office of High Energy and Nuclear Physics, Division of High 
Energy Physics of the U.S. Department of Energy under Contract 
DE-AC03-76SF00098.

\end{document}